\documentclass[final,5p,times,twocolumn]{elsarticle}

\usepackage{amssymb}

\journal{Physica E}

\usepackage{color}

\begin{document}

\begin{frontmatter}

\title{Polarized superfluid state 
in a three-dimensional fermionic optical lattice}

\author[1]{A. Koga}
\author[2]{J. Bauer}
\author[3]{P. Werner}
\author[4]{Th. Pruschke}

\address[1]{Department of Physics, Tokyo Institute of Technology, Tokyo 152-8551, Japan}
\address[2]{Max-Planck Institute for Solid State Research, Heisenbergstr. 1, 70569 Stuttgart, Germany}
\address[3]{Theoretische Physik, ETH Zurich, 8093 Z\"urich, Switzerland}
\address[4]{Institut f\"ur Theoretische Physik Universit\"at G\"ottingen,
G\"ottingen D-37077, Germany}

\begin{abstract}
We study ultracold fermionic atoms trapped 
in a three dimensional optical lattice 
by combining the real-space dynamical mean-field approach with continuous-time 
quantum Monte Carlo simulations. 
For a spin-unpolarized system we show results the density and
pair potential profile in the trap for a range of temperatures.
We discuss how a polarized superfluid state is spatially realized 
in the spin-polarized system
with harmonic confinement at low temperatures and present the local particle
density, local magnetization, and pair potential.
\end{abstract}


\end{frontmatter}



\section{Introduction}\label{s1}

Since the successful realization of Bose-Einstein condensation 
in bosonic $\rm ^{87}Rb$ \cite{Rb} and $\rm ^{23}Na$ \cite{Na} systems, 
ultracold atoms have attracted considerable interest~\cite{Review,PethickSmith,Pitaevskii}.
One of the most active research areas in this field 
concerns optical lattice systems, 
which are generated by subjecting the trapped ultracold atoms 
to a periodic potential generated by appropriated laser beams~\cite{BlochGreiner,Bloch,Jaksch,Morsch}.
The setups provide clean quantum systems with parameters 
which can be tuned in a controlled fashion.
Remarkable phenomena have been observed 
such as the phase transition between a Mott insulator and a superfluid 
in bosonic systems~\cite{Greiner}. 
In the fermionic case,
both the superfluid state~\cite{Chin} and 
the Mott insulating state~\cite{Joerdens,Schneider} have been observed.
Furthermore, spin imbalanced populations
have recently been realized~\cite{Imbalance1,Imbalance2},
which stimulate further theoretical and experimental
investigations on ultracold fermionic systems, and allow to investigate
well-known ideas from conventional condensed matter physics.

For the imbalanced system with attractive interactions,
interesting ordered ground states have been proposed as the standard s-wave
pairing at the Fermi surface is then modified.
One of the most prominent candidates is 
the Fulde-Ferrell-Larkin-Ovchinnikov (FFLO) 
phase \cite{FFLO1,FFLO2},
in which Cooper pairs with nonzero total momentum are formed.
This phase has been observed in the high field region 
in ${\rm CsCoIn_5}$ \cite{CsCoIn1,CsCoIn2,CsCoIn3},
and has theoretically been discussed in this compounds \cite{FFLOH}, 
as well as cold atoms with imbalanced populations \cite{FFLOO,FFLO1D}.
Another proposed phase is 
the breached-pair (BP) phase, where
both the superfluid order parameter and the magnetization are finite 
at zero temperature \cite{BP1,BP2,BP3,BP4,BP5}.
When one considers three dimensional optical lattice systems
at finite temperatures, 
the naively expected polarized superfluid state may be more stable than
the others. 
However, it is not clear how the polarized superfluid state is realized,
and how the pair potential and the magnetization are spatially distributed
in the imbalanced system with a confining potential.
Understanding this issue may be important to observe 
the polarized superfluid state experimentally.

In order to clarify these aspects, we investigate the attractive Hubbard model
with imbalanced spin populations and a confining potential. This allows to
discuss the effect of the imbalanced spin populations on the superfluid state. 
By using real-space dynamical mean-field theory (R-DMFT) 
\cite{Metzner,Muller,Georges,Pruschke},
we study the low temperature properties of the model.
Here, we use the continuous-time quantum Monte Carlo (CTQMC) method  \cite{Rubtsov}
based on the Nambu formalism as an impurity solver \cite{KogaWerner}.
By calculating the local particle density, local magnetization
and  pair potential, we clarify how the polarized superfluid state 
is realized in the spin imbalanced system within the confining potential.

The paper is organized as follows. 
In Sec. \ref{2}, we introduce the model Hamiltonian and summarize
various aspects of the R-DMFT. 
We demonstrate how the superfluid state is realized 
in a fermionic optical lattice with the confining potential in Sec. \ref{3}.
A brief summary is given in the last section.

\section{Model Hamiltonian and Method}\label{2}
Let us consider ultracold fermionic atoms in an optical lattice
with harmonic confinement, 
which may be described by the following attractive Hubbard Hamiltonian
~\cite{Scalettar,Micnas,Freericks,Keller,Capone,Garg,Toschi,Bauer},
\begin{eqnarray}
H&=&\sum_{\langle ij\rangle \sigma}-t c_{i\sigma}^\dag c_{j\sigma}
+\sum_{i \sigma}\left[-\left(\mu+h\sigma\right)+
V\left(\frac{r_i}{a}\right)^2\right] n_{i\sigma}\nonumber\\
&-&U\sum_i \left[n_{i\uparrow}n_{i\downarrow}-\frac{1}{2}
\left(n_{i\uparrow}+n_{i\downarrow}-1\right)\right],
\end{eqnarray}
where $c_{i\sigma} (c_{i\sigma}^\dag)$ annihilates (creates) 
a fermion at the $i$th site with spin $\sigma$ and 
$n_{i\sigma} = c_{i\sigma}^\dag c_{i\sigma}$.
$h$ acts as a magnetic field which allows as to tune the spin population
imbalance, $\mu$ is the chemical potential, 
$t(>0)$ denotes the nearest neighbor hopping and 
$U(>0)$ the attractive interaction. 
$V$ is the curvature of the harmonic potential. 
The notation $\langle ij \rangle$ indicates that 
the sum is restricted to nearest neighbors.
$r_i$ is the distance measured from the center of the trap and 
$a$ is the lattice spacing.

The ground-state properties of the Hubbard model 
on inhomogeneous lattices have been studied 
theoretically  by various methods such as 
the Bogoljubov-de Gennes (BdG) equations~\cite{Chen}, 
the Gutzwiller approximation~\cite{Yamashita},
the slave-boson mean-field approach~\cite{Ruegg}, 
variational Monte Carlo simulations~\cite{Fujihara},
and the local density approximation\cite{Dao}.
On the other hand, there are few studies addressing the effect of
imbalanced populations beyond the static mean-field approach.
The density matrix renormalization group 
method\cite{Machida,Xianlong} and 
the quantum Monte Carlo method\cite{Rigol,Pour} 
are powerful for low dimensional systems, 
but it may encounter difficulties  
when applied it to higher dimensional systems.
Here we use the R-DMFT approach~\cite{Metzner,Muller,Georges,Pruschke},
where local particle correlations are taken into account precisely. 
This treatment is formally exact for the homogeneous lattice model 
in infinite dimensions \cite{Metzner,Muller,Georges,Pruschke} and
the method has successfully been applied to some inhomogeneous 
correlated systems such as the surface~\cite{Potthoff} or 
the interface of a Mott insulators~\cite{Okamoto}, 
and to fermionic atoms~\cite{Helmes,Snoek,Koga}. 

In R-DMFT, the lattice model is mapped to a collection of effective impurity 
models.
The lattice Green function is then obtained via a self-consistency
condition imposed on the impurity problems.
When one describes the superfluid state in the framework of R-DMFT \cite{Metzner,Muller,Georges,Pruschke},
the lattice Green's function should be represented in the Nambu-Gor'kov formalism. 
For a system with $L$ lattice sites it is then given by an $(L\times L)$
matrix, where each component consists of  a $(2\times 2)$ matrix as,
\begin{eqnarray}
\left[\hat{G}_{lat}^{-1}(i\omega_n)\right]_{ij}&=&
-t\delta_{\langle ij \rangle}\hat{\sigma}_z\nonumber\\
&+&\delta_{ij}\left[ \left(i\omega_n+h\right) \hat{\sigma}_0 
+ \mu_i \hat{\sigma}_z
-\hat{\Sigma}_i (i\omega_n)\right],
\label{eq:lat}\end{eqnarray}
where $i, j= 1, 2, \cdots, L$, $\mu_i = \mu - V (r_i/a ) ^2$,
$\hat{\bf \sigma}_z$ is 
the $z$ component of the Pauli matrix, 
$\hat{\sigma}_0$ the identity matrix, 
$\omega_n = (2n+1)\pi T$ the Matsubara frequency, and 
$T$ the temperature. $\delta_{\langle ij \rangle}$ is 1 when site $i$ and $j$
are neighboring sites and zero otherwise.
The site-diagonal self-energy at the $i$th site is given 
by the following $(2\times 2)$ matrix,
\begin{equation}
\hat{\Sigma}_i\left(i\omega_n\right)=\left(
\begin{array}{cc}
\Sigma_{i\uparrow}\left(i\omega_n\right) & S_i\left(i\omega_n\right)\\
S_i\left(i\omega_n\right) & -\Sigma_{i\downarrow}^*\left(i\omega_n\right)
\end{array}
\right),
\end{equation}
where $\Sigma_\sigma(i\omega_n) \; [S(i\omega_n)]$ 
is the normal (anomalous) part of the self-energy. 
In R-DMFT, the self-consistency condition is given by 
\begin{eqnarray}
\hat{G}_{lat,ii}\left(i\omega_n\right) &=& \hat{G}_{imp,i}\left(i\omega_n\right),
\end{eqnarray}
where ${\hat G}_{imp,i}$ is the Green's function 
of the effective impurity model for the $i$th site.
Then the effective medium for each site is given by 
\begin{eqnarray}
\hat{\cal G}^{-1}_i\left(i\omega_n\right) &=&\left[\hat{G}_{imp,i}
\left(i\omega_n\right)\right]^{-1}+
\hat{\Sigma}_i\left(i\omega_n\right).
\end{eqnarray}
In this paper, we focus on the low energy state 
without lattice symmetry breaking\cite{Koga}.
In this case, the point group symmetry can be employed to efficiently  
deduce the lattice Green's function.
The inverse lattice Green's function eq. (\ref{eq:lat}) is  
transformed in terms of a unitary matrix $U$, as
\begin{eqnarray}
M=U {\hat G}_{lat}^{-1} U^{-1} &=& 
\left(
\begin{array}{cccc}
M_{A_{1g}} &0&\cdots&0\\
0&M_{A_{2g}} &\cdots&0\\
\vdots&\vdots&\ddots&0\\
0&0&0&M_{T_{2u}}
\end{array}
\right),
\end{eqnarray}
where $M_i$ is the $(m_i\times m_i)$ matrix,  $m_i (\le L)$ is 
the number of elements, and $i$ runs over the representations of the group $O_h$.
The lattice Green's function is then obtained by
\begin{eqnarray}
{\hat G}_{lat} &=& \left(U^{-1} M U\right)^{-1} \nonumber\\
&=& U^{-1} \left(
\begin{array}{cccc}
\left[M_{A_{1g}}\right]^{-1} &0&\cdots&0\\
0&\left[M_{A_{2g}}\right]^{-1} &\cdots&0\\
\vdots&\vdots&\ddots&0\\
0&0&0&\left[M_{T_{2u}}\right]^{-1}
\end{array}
\right) U,\nonumber\\
&&\\
{\hat G}_{lat,ii} &=& \sum_k \sum_{m,n}^{m_k} 
U_{(km),i} \; [M_k^{-1}]_{mn} \; U_{(kn),i}\label{eqlat}
\end{eqnarray}
When one considers a system size $r_i \le 7a$,
the total number of sites $L=1419$. 
In this case, there are $58$ inequivalent sites. 
Therefore, by solving fifty-eight kinds of effective impurity models 
and making use of eq. (\ref{eqlat}), we can iteratively solve the R-DMFT equations 
and discuss the stability of the polarized superfluid state
in the large cluster.

When R-DMFT is applied to our inhomogeneous system, 
it is necessary to solve a large number of effective impurity models.
There are various numerical techniques such as 
exact diagonalization\cite{Caffarel} and
the numerical renormalization group \cite{Bauer,NRG,NRG_RMP,OSakai}.
One of the most powerful methods is CTQMC, which has been developed recently.
In this method, Monte Carlo samplings of collections of diagrams 
for the partition function
are performed in continuous time.
Therefore, the Trotter error, which originates from the Suzuki-Trotter
decomposition, is avoided. 
Furthermore, this method is efficient and applicable 
to more general classes of models 
than, for example, the Hirsch-Fye algorithm\cite{Hirsch}.
The CTQMC method has successfully been applied to 
various systems such as the Hubbard model \cite{KogaWerner,CTQMC,Multi}, 
the periodic Anderson model \cite{Luitz},
the Kondo lattice model \cite{Otsuki},
and the Holstein-Hubbard model.\cite{Phonon}
Here, we use the continuous-time auxiliary field (CTAUX) version 
of weak-coupling CTQMC \cite{KogaWerner,CTAUX}
to discuss the superfluid state in the optical lattice system.
Some details of the CTQMC method are explained in \ref{appendix}.

\section{Results}\label{3} 

\begin{figure}[htb]
\begin{center}
\includegraphics[width=8cm]{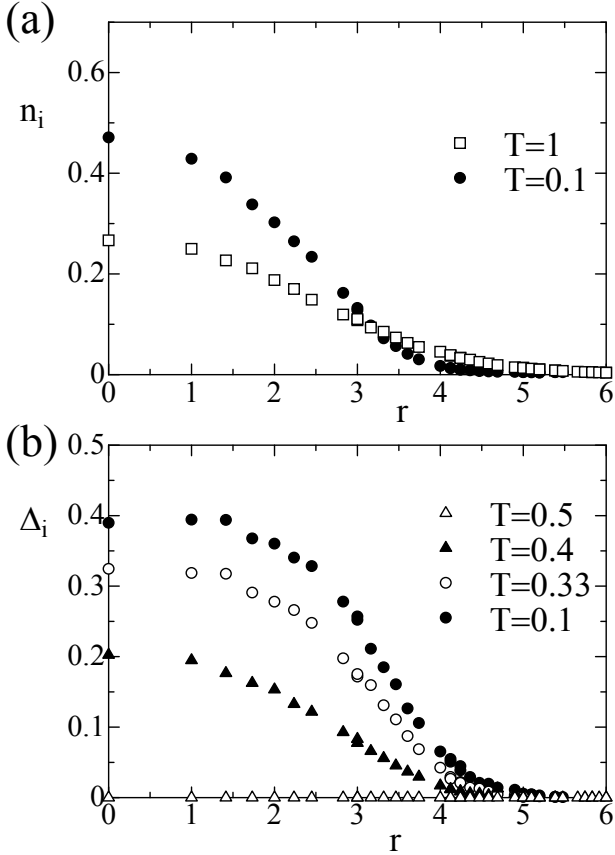}
\caption{
Profiles of particle density $\langle n_{i\sigma}\rangle$ and 
pair potential $\Delta_i$ as a function of $r$ when $V_0=0.1$, $U=8$ and 
$N_\uparrow = N_\downarrow = 40$.
}
\label{fig1}
\end{center}
\end{figure}

We now consider a three-dimensional optical lattice system 
with a confining potential to discuss how the superfluid state 
is realized in this spatially inhomogeneous set-up.
In this paper, we use the hopping integral $t=1$ as the unit of the energy and
fix the confining potential as $V=0.1$ and the total particle number as $N=80$.
We calculate site-dependent static physical quantities such as 
the particle density $n_i$, the pair potential $\Delta_i$ and 
the magnetization $m_i$, which are defined by
\begin{eqnarray}
n_i &=& \sum_\sigma \langle c_{i\sigma}^\dag c_{i\sigma} \rangle = 2-
\sum_\sigma G_{i\sigma}(0_+)\\
\Delta_i&=&\langle c_{i\uparrow} c_{i\downarrow} \rangle=F_i(0_+),\\
m_i&=&\sum_\sigma \sigma\langle c_{i\sigma}^\dag c_{i\sigma} \rangle 
= -\sum_\sigma \sigma G_{i\sigma}(0_+).
\end{eqnarray}
By performing R-DMFT with the CTQMC method, 
we obtain these quantities first for the balanced system
$(N_\uparrow = N_\downarrow = 40)$, as shown in Fig. \ref{fig1}.
At high temperatures, 
particle correlations are small and kinetic energies high such that the 
fermions are widely distributed in the trap, 
as shown in Fig. \ref{fig1} (a).
Decreasing the temperature, the particles gather around the center of the system,
which is due to the existence of the attractive interaction.
Below a certain critical temperature $T_c = 0.4 \sim 0.5$, 
the pair potential $\Delta_i$ becomes finite,
as shown in Fig. \ref{fig1} (b). 
This means that a superfluid state is realized in the region with $n_i \neq 0$.
It is also found that the magnitude of pair potential depends on the site.
This originates from the existence of the harmonic confinement in the system,  
which is consistent with the results obtained from BdG equations \cite{Chen}.

\begin{figure}[htb]
\begin{center}
\includegraphics[width=8cm]{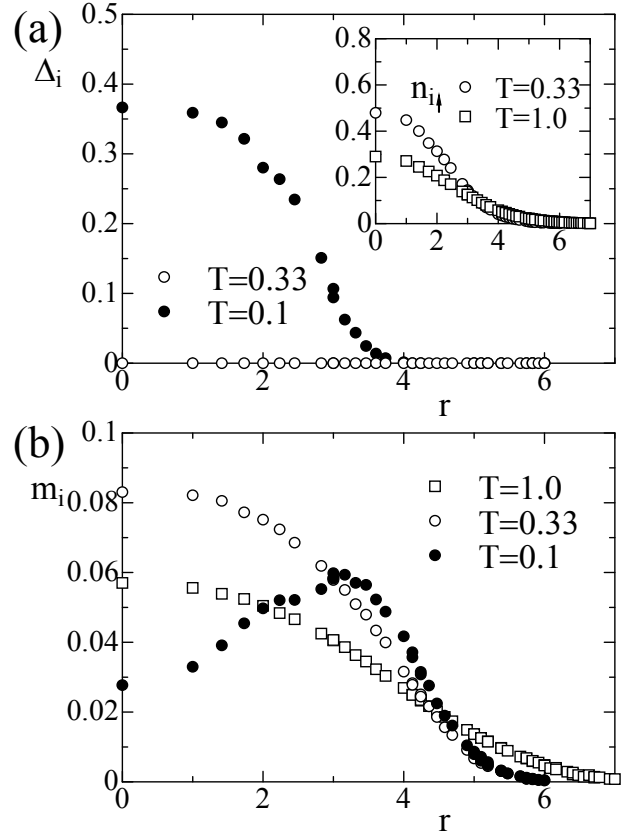}
\caption{
Profiles of pair potential $\Delta_i$ and local magnetization $m_i$ 
as a function of $r$
when $V_0=0.1$, $U=8$, $N_\uparrow = 50$ and $N_\downarrow = 30$. 
Inset of (a) shows the profile of particle density 
$\langle n_{i\uparrow}\rangle$.
}
\label{fig2}
\end{center}
\end{figure}

We next discuss the effect of the imbalanced populations.
The obtained results for $P =  0.25$ are shown in Fig. \ref{fig2}, 
where $P = (N_\uparrow - N_\downarrow)/(N_\uparrow + N_\downarrow)$ is 
the spin imbalance parameter.
This case is similar to a system with applied 
magnetic field.
Therefore, the superfluid state should become unstable \cite{KogaWerner}.
In fact, in contrast to the balanced system, 
the normal metallic state is stabilized 
even at $T = 0.33$.
Further decrease of the temperature $(T=0.1)$ 
induces the pair potential in a certain region $(r_i < 4a)$,
as shown in Fig. \ref{fig2} (a).
Since the magnetization is finite in this region, we conclude that 
the polarized superfluid state is realized in the region $(0<r_i<4a)$.
On the other hand, the normal metallic state still remains
in the region $(4a<r_i<6a)$.
Note that at $T=0.1$, the pair potential has a maximum in the center of the
system, where particle pairs are strongly formed.
As can be seen in Fig. \ref{fig2} (b), the magnetization at the center of the
trap first increases on reducing the temperature. However, when the system
becomes superfluid, the magnetization is suppressed and the maximum is pushed
to larger values of $r_i$ resulting in a non-monotonic behavior appears in the
magnetization curve. 
It is expected that further decrease of the temperature will lead to a complete
suppression of the magnetization near the center of the trap.
Detailed calculations will be presented elsewhere.

\section{Summary}\label{4}
We have investigated ultracold fermionic atoms trapped in an optical lattice 
with spin imbalanced populations.
By combining the real-space 
dynamical mean-field theory with continuous-time quantum 
Monte Carlo simulations 
based on the Nambu formalism,
we have calculated the local particle density, local magnetization, and 
the pair potential in the system.
We have demonstrated how a polarized superfluid state is spatially realized 
at low temperatures in the model with harmonic confinement.

In this paper, we have studied the stability of 
the polarized superfluid state in a system
with small number of fermions. 
If the total particle number is larger, 
the local particle density may increase beyond $n_i = 0.5$.
It is known that the density wave state and the superfluid state
are degenerate in the homogeneous system at half filling $(n=0.5)$
except in one dimension, which means that a supersolid state 
might be realizable in an optical lattice system with 
 modulated particle density.
It is an interesting problem to clarify this
by means of our method. Work along those lines is in progress.

\section*{Acknowledgment}
The authors thank N. Kawakami for valuable discussions. 
Parts of the computations were done on TSUBAME
Grid Cluster at the Global Scientific Information and Computing
Center of the Tokyo Institute of Technology. 
This work was partly supported by the Grant-in-Aid for Scientific Research 
20740194 (A.K.) and 
the Global COE Program ``Nanoscience and Quantum Physics" from 
the Ministry of Education, Culture, Sports, Science and Technology (MEXT) 
of Japan. PW acknowledges support from SNF Grant PP002-118866.

\appendix
\section{Continuous-Time Quantum Monte Carlo simulations in the Nambu Formalism}
\label{appendix}

This appendix explains some details of the CTAUX method \cite{CTAUX}.
When the superfluid state is discussed in the framework of DMFT,
the total particle number is not conserved 
in the effective Anderson impurity model.
The Hamiltonian is then given by
\begin{eqnarray}
H&=&H_0 + H_U,\label{eq:anderson}\\
H_0&=&\sum_{p\sigma} \epsilon_{p\sigma} n_{p\sigma}
+\sum_{p\sigma}\left(V_{p\sigma} d_\sigma^\dag a_{p\sigma}+h.c.\right)
\nonumber\\
&+&\sum_{p}\left(\Delta_p a_{p\uparrow}^\dag a_{p\downarrow}^\dag +h.c.\right)
+\sum_\sigma E_{d\sigma} n_{d\sigma},\\
H_U&=&-U\left[n_{d\uparrow} n_{d\downarrow}
-\frac{1}{2}\left(n_{d\uparrow}+n_{d\downarrow}-1\right)\right],
\end{eqnarray}
where $a_{p\sigma}$ $(d_\sigma)$ annihilates a fermion with spin $\sigma$ 
in the $p$th orbital of the effective baths (the impurity site).
The effective bath is represented by $\epsilon_{p\sigma}$ and $\Delta_p$,
and $V_{p\sigma}$ represents the hybridization 
between the effective bath and the impurity site. 
$E_{d\sigma}$ is the energy level for the impurity site,
$n_{p\sigma}=a_{p\sigma}^\dag a_{p\sigma}$, and
$n_{d\sigma}=d_\sigma^\dag d_\sigma$. 
The Green's function should be defined by the $2\times 2$ matrix, as
\begin{eqnarray}
\hat{G}(\tau)  
&=&\left(
\begin{array}{cc}
G_\uparrow(\tau)&F(\tau)\\
F^*(\tau)&-G_\downarrow(-\tau)
\end{array}
\right),
\end{eqnarray}
where
\begin{eqnarray}
G_\sigma(\tau) &=& \langle T_\tau c_\sigma(\tau)
c_\sigma^\dag(0)\rangle,\\
F(\tau) &=& \langle T_\tau c_\uparrow(\tau) c_\downarrow(0)\rangle,\\
F^*(\tau) &=& \langle T_\tau c^\dag_\downarrow(\tau)
c^\dag_\uparrow(0)\rangle,
\end{eqnarray}
where $T_\tau$ is the imaginary-time ordering operator and 
we have chosen the Green's functions $G_\sigma(\tau)$ to be positive.

To perform simulations, 
we consider here a weak coupling CTQMC approach.
The partition function $Z$ is given by
\begin{eqnarray}
Z&=&{\rm Tr}\left[e^{-\beta H_1} T_\tau e^{-\int_0^\beta d\tau H_2(\tau)} 
\right]\nonumber\\
&=& \sum_{n=0}^\infty \int_0^\beta d\tau_1 \int_{\tau_1}^\beta d\tau_2 \cdots \int_{\tau_{n-1}}^\beta d\tau_n \nonumber\\
&\times&(-1)^n {\rm Tr}\left[
e^{-\beta H_1}H_2(\tau_n)H_2(\tau_{n-1})\cdots H_2(\tau_1)
\right],
\end{eqnarray}
where $H_2(\tau) = e^{\tau H_1}H_2 e^{-\tau H_1}$ and $\beta = 1/T$.
Here, we have divided the impurity Hamiltonian Eq.~(\ref{eq:anderson})
into two parts as,
\begin{eqnarray}
H_1&=&H-H_2,\\
H_2&=&H_U-K/\beta\nonumber\\
&=&\frac{K}{2\beta}\sum_{s=-1,1} e^{\gamma s \left(n_\uparrow+n_\downarrow-1\right)},
\end{eqnarray}
with $\gamma= \cosh^{-1} (1+\beta U/2 K)$, and $K$ some nonzero constant.
In this paper, we set $K=1$ in the CTQMC simulations.
The introduction of the Ising variable $s$ in $H_2$ 
enables us to perform simulations at arbitrary filling. 
An $n$th order configuration
$c = \{ s_1, s_2, \cdots , s_n; \tau_1, \tau_2, \cdots, \tau_n\}$
corresponding to auxiliary spins $s_1,s_2,\ldots,s_n$ at imaginary times $\tau_1<\tau_2<\ldots<\tau_n$
contributes a weight
\begin{eqnarray}
w_c &=& e^{-K}\left(\frac{Kd\tau}{2\beta}\right)^n
e^{-\gamma \sum s_i}Z_0\;{\rm det}\; 
\left[\hat{N}^{(n)}\right]^{-1}
\end{eqnarray}
to the partition function.
Here, $Z_0={\rm Tr}[e^{-\beta H_1}]$ and $\hat{N}^{(n)}$ is an $n\times n$ matrix, 
where each element consists of a $2\times 2$ matrix:
\begin{eqnarray}
\left[\hat{N}^{(n)}\right]^{-1} &=& \hat{\Gamma}^{(n)}
-\hat{g}^{(n)}\left(\hat{\Gamma}^{(n)}-\hat{I}^{(n)}\right),\\
\hat{I}^{(n)}_{ij}&=&\delta_{ij}\hat{\sigma}_0,\\ 
\hat{\Gamma}^{(n)}_{ij} &=& \delta_{ij} e^{\gamma s_i}\hat{\sigma}_0,\\
\hat{g}^{(n)}_{ij} &=&\left(
\begin{array}{cc}
g_{0\uparrow}(\tau_i-\tau_j)&f_0(\tau_i-\tau_j)\\
-f_0^*(\tau_i-\tau_j)&g_{0\downarrow}(\tau_j-\tau_i)
\end{array}
\right),\hspace{4mm}
\end{eqnarray}
with $i,j = 1,2, \cdots n$.
The sampling process must satisfy
ergodicity and (as a sufficient condition) detailed balance.
For ergodicity, it is enough to insert or remove the Ising variables
with random orientations at random times to generate all possible configurations.

To satisfy the detailed balance condition, we decompose the transition probability as
\begin{eqnarray}
p\left(i\rightarrow j\right) = p^{\rm prop}\left(i\rightarrow
j\right)p^{\rm acc}\left(i\rightarrow j\right),
\end{eqnarray}
where $p^{\rm prop} (p^{\rm acc})$ is the probability to propose (accept) 
the transition from the configuration $i$ to the configuration $j$. 
Here, we consider the insertion and removal of the Ising spins 
as one step of the simulation process, 
which corresponds to a change of $\pm 1$ in the perturbation order.
The probability of insertion/removal of an Ising spin is then given by 
\begin{eqnarray}
p^{\rm prop}(n\rightarrow n+1)&=&\frac{d\tau}{2\beta},\\
p^{\rm prop}(n+1\rightarrow n)&=&\frac{1}{n+1}.
\end{eqnarray}
For this choice, the ratio of the acceptance probabilities becomes
\begin{eqnarray}
\frac{p^{\rm acc}\left(n\rightarrow n+1\right)}
{p^{\rm acc}\left(n+1\rightarrow n\right)}
= \frac{K}{n+1}e^{-\gamma s_{n+1}}\frac{\det N^{(n)}}
{\det N^{(n+1)}}.
\end{eqnarray}
When the Metropolis algorithm is used to sample the configurations, 
we accept the transition 
from $n$ to $n\pm1$ with the probability
\begin{eqnarray}
{\rm min}\left[1, \frac{p^{\rm acc}\left(n\rightarrow n\pm 1\right)}
{p^{\rm acc}\left(n\pm 1\rightarrow n\right)}\right].
\end{eqnarray}

In each Monte Carlo step, we measure the following Green's functions 
($0<\tau<\beta$),
\begin{eqnarray}
G_\sigma(\tau)&=&\frac{1}{Z}{\rm Tr}
\left[e^{-\beta H}c_\sigma(\tau)c^\dag_\sigma(0)\right],\\
F(\tau)&=&\frac{1}{Z}{\rm Tr}
\left[e^{-\beta H}c_\uparrow(\tau)c_\downarrow(0)\right],\\
F^*(\tau)&=&\frac{1}{Z}{\rm Tr}
\left[e^{-\beta H}c^\dag_\downarrow(\tau)c^\dag_\uparrow(0)\right].
\end{eqnarray}
By using Wick's theorem, the contribution of a certain configuration $c$ is given by 
\begin{eqnarray}
G_{\sigma}^{c}(\tau) &= \det [N^{(n)}] 
{\rm det}\left(
\begin{array}{cc}
\left[N^{(n)}\right]^{-1}&Q_\sigma\\
R_\sigma&g_{0\sigma}(\tau)
\end{array}
\right),\\
F^{c}(\tau) &= \det [N^{(n)}] 
{\rm det}\left(
\begin{array}{cc}
\left[N^{(n)}\right]^{-1}&Q'\\
R'& f_0(\tau)
\end{array}
\right),\\
F^{*c}(\tau) &= \det [N^{(n)}] 
{\rm det}\left(
\begin{array}{cc}
\left[N^{(n)}\right]^{-1}&Q^{*\prime}\\
R^{*\prime}& f^{*}_0(\tau)
\end{array}
\right),
\end{eqnarray}
where $Q_\sigma, Q^{\prime}, Q^{*\prime}, R_\sigma, R^\prime, R^{*\prime}$ are vectors,
in which the $i$th element ($i=1,2,\cdots, n$) is defined by
\begin{eqnarray}
Q_{\uparrow i} &=& \{-g_{0\uparrow}(\tau_i)\;\; f^*_0(\tau_i)\}^T,\\
Q_{\downarrow i} &=& \{f_0(\tau_i-\tau)\;\; g_{0\downarrow}(\tau-\tau_i)\}^T,\\
Q'_i &=& \{-f_0(\tau_i)\;\; -g_{0\downarrow}(-\tau_i)\}^T,\\
Q^{*\prime}_i &=& Q_{\uparrow i},\\
R_{\uparrow i}&=& (e^{\gamma s_i}-1)\{g_{0\uparrow}(\tau-\tau_i)\;\; 
f_0(\tau-\tau_i)\},\\
R_{\downarrow i}&=& (e^{\gamma s_i}-1)\{f^*_0(-\tau_i)\;\;
-g_{0\downarrow}(\tau_i)\},\\
R'_i &=& R_{\uparrow i},\\
R^{*\prime}_i &=& (e^{\gamma s_i}-1)\{f^*_0(\tau-\tau_i)\;\;
-g_{0\downarrow}(\tau_i-\tau)\}.\hspace{2mm}
\end{eqnarray}

\end{document}